\documentclass[superscriptaddress,twocolumn,showpacs,amsmath,amssymb]{revtex4}
\usepackage{graphicx}
\usepackage{color}

\begin{document}

\title{
Enhancement factor for two-neutron transfer reactions with a schematic 
coupled-channels model}

\author{K. Hagino}
\affiliation{
Department of Physics, Tohoku University, Sendai 980-8578,  Japan}
\affiliation{Research Center for Electron Photon Science, Tohoku
University, 1-2-1 Mikamine, Sendai 982-0826, Japan}
\affiliation{
National Astronomical Observatory of Japan, 2-21-1 Osawa,
Mitaka, Tokyo 181-8588, Japan}

\author{G. Scamps}
\affiliation{
Department of Physics, Tohoku University, Sendai 980-8578,  Japan}


\begin{abstract}
Probabilities for two-neutron transfer reactions, $P_{\rm 2n}$, 
are often discussed in comparison with the square of the 
corresponding probabilities for 
one-neutron transfer process, $(P_{\rm 1n})^2$, implicitly assuming that 
$(P_{\rm 1n})^2$ provides the probability of two-neutron transfer 
reactions in the absence of the pairing correlation. 
We use a schematic coupled-channels model, in which the transfers 
are treated as effective inelastic channels, and demonstrate 
that this model leads to $P_{\rm 2n}=(P_{\rm 1n})^2/4$, 
rather than $P_{\rm 2n}=(P_{\rm 1n})^2$, 
in the pure sequential limit. 
We argue that a simple model with spin-up and spin-down neutrons in a 
single-particle orbital also leads to the same conclusion. 
\end{abstract}


\maketitle

It has been well known that the pairing correlation enhances 
cross sections for the two-neutron transfer process as compared to 
those in the uncorrelated 
limit \cite{Yoshida62,BW91,OV01,VS12,Potel13,LFV14}. 
Those cross sections 
are often converted to the transfer probabilities by dividing 
them by the Rutherford cross sections, and are 
plotted as a function of the 
distance of the closest approach, $D$, for the classical Rutherford 
trajectory. This representation in fact provides a convenient way to 
discuss the reaction dynamics, since the cross sections for different 
values of incident energies and the scattering angles can be analysed in 
a unified way. 
The enhancement of the two-neutron transfer process has customary been 
discussed by taking the ratio between $P_{\rm 2n}$ 
and $(P_{\rm 1n})^2$ 
\cite{OV01,vonOertzen83,vonOertzen87,Wu90,Liu91,Peter99,CSP11,MCS14,CPS09}, 
where $P_{\rm 1n}$ and $P_{\rm 2n}$ are the probabilities for the one- and 
two-neutron transfer processes, respectively. 
That is, it has been usually believed that the quantity 
$(P_{\rm 1n})^2$ provides a reference probability for the two-neutron 
transfer process which would be realized in the absence of the pairing 
correlation \cite{vonOertzen91}. 

In this paper, we discuss 
the validity of this assumption. To this end, we consider 
the two-neutron transfer probability 
in the no-correlation limit, where the two-neutron transfer process 
takes place in a completely sequential manner. 
This work is partly motivated by a recent result of a time-dependent 
Hartree-Fock (TDHF) + BCS calculation, which shows that the ratio 
$P_{\rm 2n}/(P_{\rm 1n})^2$ in the absence of the pairing correlation is 
well parametrized as \cite{SL13}, 
\begin{equation}
\frac{P_{\rm 2n}}{(P_{\rm 1n})^2}\sim \frac{N_v-1}{2N_v}\cdot 
\frac{N_f-1}{N_f},
\label{TDHF}
\end{equation}
where $N_v$ and $N_f$ are the number of valence nucleons and the 
number of available states in the receiver nucleus, respectively. 
This equation 
suggests that the ratio $P_{\rm 2n}/(P_{\rm 1n})^2$ is not unity in 
general, but is more complex and never exceeds 1/2. 
In this paper, we 
employ the coupled-channels approach 
to investigate this problem from a different perspective. 
In particular, we use a schematic coupled-channels model for two-neutron 
transfer, and attempt to understand the result of TDHF. 

In the coupled-channels approach to transfer reactions, one often 
treats transfer channels as effective inelastic 
excitations \cite{LPE88,Esb89,Esb98,Rowley01}. 
In this paper, we use the same treatment 
for the transfer channels 
and consider the following 
coupling matrix for a sequential two-neutron transfer 
reaction \cite{Rowley92,SH15}, 
\begin{equation}
V=
    \left(
    \begin{array}{ccc}
    0 & F(r) & 0 \\
    F(r) & -Q & F(r)\\
    0 & F(r)&  -2Q
    \end{array}
    \right). 
\label{seq}
\end{equation}
Here, we have assumed that all the channels have zero angular momentum. 
In this equation, $F(r)$ is the form factor for the coupling between the 
entrance (0n) and the one-neutron (1n) transfer 
channels, while $Q$ is the $Q$-value 
for the 1n-transfer reaction. 
In this coupling scheme, the 0n channel is coupled to the 1n channel, which 
is sequentially coupled to the two-neutron (2n) channel. 
The no-correlation limit is simulated by 
setting the coupling between the 1n and the 2n transfer channels to be 
the same as that between the 0n and the 1n transfer channels, and also by 
setting the $Q$-value for the 2n transfer channel to be exactly 
twice the $Q$-value 
for the 1n channel. 
The direct coupling between the 0n and the 2n channels is also set to 
be zero. 

We apply this model to the $^{40}$Ca+$^{96}$Zr reaction, 
for which the experimental transfer cross sections have been reported in 
Ref. \cite{CSP11}. To this end, we use 
a function which asymptotically has an exponential form, 
\begin{equation}
F(r)\sim \frac{\beta}{a} e^{-(r-R)/a},
\label{formfactor}
\end{equation}
for the coupling form 
factor $F(r)$, and set the transfer $Q$ value to be $Q$=0 \cite{Rowley01,SH15}. 
(In the actual calculations shown below, for a numerical reason, 
we use a derivative form of the Fermi function with the parameters $\beta, R$, and $a$.)
With the Woods-Saxon type for the nuclear potential, with the parameters 
of $V_0$ = 140 MeV, $r_0$ = 1.1 fm, and $a_0$ = 0.65 fm for the real part and 
$W_0$ = 30 MeV, $r_W$ = 1.15 fm and $a_W$ = 0.1 fm for the imaginary part, we 
vary the parameters in the coupling form factor, Eq. (\ref{formfactor}), 
so that the experimental data for the one-neutron transfer reaction 
can be reproduced. 
To this end, the coupled-channels 
equations are solved 
using a version of the computer code {\tt CCFULL} \cite{HRK99}. 
The resultant values for the parameters 
are $\beta$ = 9 MeV fm, $R$ = 1.15 $\times (40^{1/3} 
+96^{1/3})$ fm, and $a$ = 1.3 fm. 

\begin{figure}[tb]
\includegraphics[clip,width=7.5cm]{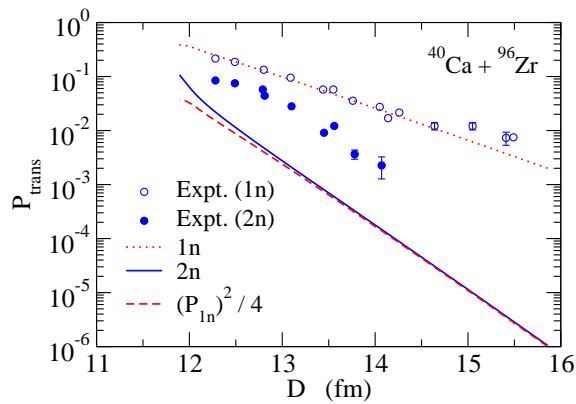}
\caption{(Color online) 
The transfer probabilities, defined as the ratio of the transfer 
cross sections to the Rutherford cross sections, 
for the $^{40}$Ca + $^{96}$Zr reaction. These probabilities 
are plotted as a function of the 
distance of the closest approach, $D$, of the classical Rutherford 
trajectory. The dotted and the solid lines denote the one- and two-neutron 
transfer probabilities, respectively, while the dashed line is 
a quarter of the square of the one-neutron transfer probability. 
The experimental data are taken from Ref. \cite{CSP11}. 
}
\end{figure}

Figure 1 shows the transfer probabilities so obtained. 
Here, the transfer probabilities are defined as the ratio 
of the transfer cross sections to the Rutherford cross sections, that is, 
$P_{xn} = (d\sigma_{xn}/d\Omega)/(d\sigma_{\rm R}/d\Omega)$, 
where $x$ = 1, 2 is the number 
of transferred neutron, $d\sigma_{xn}/d\Omega$ and $d\sigma_{\rm R}/d\Omega$ 
are the transfer and the Rutherford cross sections, respectively. 
This definition is applied both to the experimental data and 
to the theoretical 
calculations. 
The transfer probabilities are plotted as a function of the 
distance of the closest approach, $D$, of the Rutherford trajectory 
for the scattering angle of $\theta_{c.m.}$ = 140 degrees in the center 
of mass frame. The dotted and the solid lines denote the transfer 
probabilities for the 1n and the 2n channels, respectively. 
While the 1n probabilities are well reproduced, as expected, 
the 2n probabilities 
are largely underestimated by this calculation. 
One can clearly see 
that the 2n probability, $P_{\rm 2n}$, is 
consistent with a quarter of the square of the 1n probability, 
$(P_{\rm 1n})^2/4$, which is denoted by the dashed line in the figure. 

This relation can be easily understood if one uses the time-dependent 
perturbation theory. In the semi-classical coupled-channels approach, 
one assumes a classical trajectory $r(t)$ 
for the relative motion between the 
colliding nuclei, 
and solve the time-dependent coupled-channels equations 
for the intrinsic motion \cite{BW91,CPS09}. 
Applying the first and the second order perturbation theory, the amplitudes 
for the one- and the two-neutron transfer processes 
for the sequential two-neutron transfer coupling, Eq. (\ref{seq}), 
read, 
\begin{eqnarray}
a_{\rm 1n}&=&\frac{1}{i\hbar}\int^\infty_{-\infty}dt\, e^{-iQt/\hbar}F(r(t)), 
\label{1st}
\\
a_{\rm 2n}&=&
\left(\frac{1}{i\hbar}\right)^2
\int^\infty_{-\infty}dt\, e^{-iQt/\hbar}F(r(t)) \nonumber \\
&&\times \int^t_{-\infty}dt'\, 
e^{-iQt'/\hbar}F(r(t')), \\
&=&\frac{1}{2}\,\left[\frac{1}{i\hbar}
\int^\infty_{-\infty}dt\, e^{-iQt/\hbar}F(r(t))\right]^2,
\label{2nd}
\end{eqnarray}
respectively. 
The last equality is due to the property of the 
pure sequential transfer, that is, $Q_{\rm 2n}=2 Q_{\rm 1n}$ and 
$F$(1n-2n)=$F$(0n-1n). 
By squaring these equations, one obtains $P_{\rm 2n}/(P_{\rm 1n})^2 
= |a_{\rm 2n}|^2/|a_{\rm 1n}|^4 = 1/4$, which is indeed realized 
in Fig. 1 for 
large values of $D$, at which the perturbative treatment is 
justified. 

\begin{figure}[tb]
\includegraphics[clip,width=7.5cm]{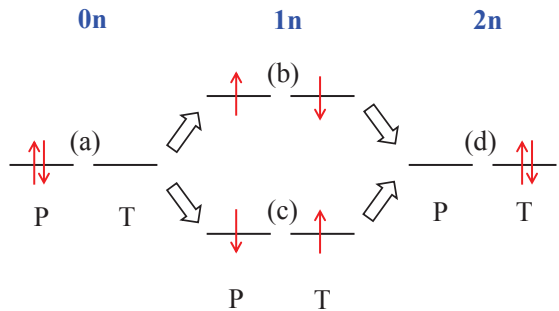}
\caption{(Color online) 
A schematic model for the two-neutron transfer process of spin-up 
and spin-down neutrons. $P$ and $T$ denote the projectile and the target 
nuclei, respectively. 
}
\end{figure}

The factor of 1/4 can also be obtained with a more microscopic 
model, which is illustrated in Fig. 2. Here we consider a transfer of 
spin-up and spin-down neutrons, which initially occupy 
a single-particle state in a projectile nucleus (see 
the state (a) in Fig. 2). One of those neutrons is initially transferred 
to a target nucleus (the states (b) or (c), depending on the spin of 
the transferred neutron), which is followed by a transfer 
of the other neutron to the target nucleus (the state (d)). 
We assume that the matrix elements for the transfer process 
do not depend on the spin of the transferred neutron, 
and that the spin flip does not occur during the transfer. 
We thus have 
$\langle \uparrow_P|V|\uparrow_T\rangle
=\langle \downarrow_P|V|\downarrow_T\rangle\equiv\tilde{F}(r)$ and 
$\langle \uparrow_P|V|\downarrow_T\rangle
=\langle \downarrow_P|V|\uparrow_T\rangle=0$, where $V$ is the operator which 
induces the transfer, and $P$ and $T$ denote the projectile and the target 
nuclei, respectively. 
We again use the time-dependent perturbation theory in order to evaluate 
the transfer probabilities. 
For the one-neutron transfer probability, 
there are two distinguishable final states, (b) and (c) in Fig. 2, 
and one has to add the probabilities 
for the processes (a)$\to$(b) and (a)$\to$(c). 
One thus obtains (see Eq. (\ref{1st})), 
\begin{equation}
P_{\rm 1n}=2\times\left|\frac{1}{i\hbar}\int^\infty_{-\infty}dt\, e^{-iQt/\hbar}
\tilde{F}(r(t))\right|^2. 
\label{1n-2}
\end{equation}
For the two-neutron transfer process, 
there are two indistinguishable paths, (a)$\to$(b)$\to$(d) and 
(a)$\to$(c)$\to$(d), to the final state, and one has to add
the amplitudes first. This leads to (see Eq. (\ref{2nd})), 
\begin{equation}
P_{\rm 2n}=\left|\frac{1}{2}\,\left[\frac{1}{i\hbar}
\int^\infty_{-\infty}dt\, e^{-iQt/\hbar}\tilde{F}(r(t))\right]^2\times 2\right|^2. 
\label{2n-2}
\end{equation}
Comparing Eq. (\ref{1n-2}) with Eq. (\ref{2n-2}), one again 
obtains $P_{\rm 2n}/(P_{\rm 1n})^2 =1/4$. It is easy to confirm that this 
relation still holds even if one considers the anti-symmetrization of 
each state, {\it e.g.,} 
$|a\rangle = (|\uparrow_P\downarrow_P\rangle 
-|\downarrow_P\uparrow_P\rangle)/\sqrt{2}$ and 
$|b\rangle = (|\uparrow_P\downarrow_T\rangle 
-|\downarrow_T\uparrow_P\rangle)/\sqrt{2}$. 

As in the multi-phonon couplings in the coupled-channels approach 
\cite{HT12,Kruppa93}, one can make a relation between the coupled-channels 
model of Eq. (\ref{seq}) and the schematic model of Fig. 2. That is, 
by introducing a single effective one-neutron transfer channel defined 
by $|1n\rangle = (|b\rangle + |c\rangle)/\sqrt{2}$, it is easy to find 
$\langle 1n|V|a\rangle = \langle d|V|1n\rangle = \sqrt{2}\tilde{F}$, 
where $|a\rangle, |b\rangle, |c\rangle$, and $|d\rangle$ are 
the states shown in Fig. 2. 
Therefore, identifying $F=\sqrt{2}\tilde{F}$, the two models are 
actually equivalent 
to each other. Notice that the other combination of the states 
$|b\rangle$ and $|c\rangle$, that is, 
$(|b\rangle - |c\rangle)/\sqrt{2}$, couples neither to $|a\rangle$ 
nor to $|d\rangle$ and is decoupled from the model space. 

The factor of 1/4 for the relation between $P_{\rm 2n}$ and 
$(P_{\rm 1n})^2$ is consistent with the previous result of TDHF, 
Eq. (\ref{TDHF}), if one disregards the dependence on $N_f$. 
Notice that the $N_f$ dependence 
in Eq. (\ref{TDHF}) was obtained by counting the number of possibilities to 
put nucleons in the final single-particle state \cite{SL13}. 
To this end, the probability 
was assumed to be 
the same for all the final states with different values of $j_z$, 
that is, the $z$-component of the single-particle angular momentum 
in the receiver nucleus. 
If one neglects the spin-flip components, however, 
the formula would become 
\begin{equation}
\frac{P_{\rm 2n}}{(P_{\rm 1n})^2}\sim \frac{N_v-1}{2N_v}, 
\end{equation}
with which one obtains 
$P_{\rm 2n}/(P_{\rm 1n})^2 =1/4$ for $N_v=2$. 

In summary, we have investigated the two-neutron transfer 
reactions in the no-correlation limit. To this end, we have used a schematic 
coupled-channels model, in which the transfer channels are treated 
as effective inelastic excitations. 
We have shown that the probability of two-neutron transfer process, $P_{\rm 2n}$, 
is given by a quarter of $(P_{\rm 1n})^2$, that is, 
$P_{\rm 2n}/(P_{\rm 1n})^2 =1/4$. This result is to some extent 
consistent with the result of 
time-dependent Hartree-Fock calculations for two valence neutrons. 
The two-neutron transfer probabilities have customary been compared with 
$(P_{\rm 1n})^2$, rather than 
$(P_{\rm 1n})^2/4$. 
Of course, many experimental data are for inclusive processes, and 
the enhancement factor for the two-neutron transfer process reflects not 
only the pairing correlation but also the phase space factor for the 
intermediate and the final states. 
Nevertheless, there is no strong reason why the two-neutron transfer 
probability should be compared with 
$(P_{\rm 1n})^2$, and we advocate using 
$(P_{\rm 1n})^2/4$, which has a clearer physical 
meaning as a reference probability, at least for 
a core+two-neutron system.

\bigskip

We thank A. Vitturi and L. Fortunato for useful discussions. 
G.S. acknowledges the Japan Society for the Promotion of Science
for the JSPS postdoctoral fellowship for foreign researchers.
This work was supported by Grant-in-Aid for JSPS Fellows No. 14F04769.

\end{document}